\documentclass[12pt]{article}
\newtheorem{theorem}{Theorem}

\newcommand{\begsection}[1]{\setcounter{equation}{0}\section{#1}}

\def\C{{\mathcal C}}

\def\R{{\mathcal R}}
\def\N{{\mathcal N}}

\def\Sc{Schr\"o\-din\-ger}   

\def\be{\begin{equation}}
\def\ee{\end{equation}}

\def\ds{\displaystyle}
\def\om{\omega}
\def\Om{\Omega}
\def\ep{\epsilon}
\def\l{\lambda}

\def\bdd{bounded}

\def\ev{eigenvalue}

\def\op{operator}

\def\Im{{\rm Im\,}}

\begin{document}
\baselineskip=21pt
\begin{center}
{\large\bf SPECTRA OF $PT-$SYMMETRIC OPERATORS AND PERTURBATION 
THEORY }
\end{center}
\vskip 13pt
\begin{center} 
Emanuela Caliceti\footnote[1]{
Dipartimento di Matematica,  Universit\`{a} di Bologna, I- 40127 
Bologna (Italy) 
 (caliceti@dm.unibo.it)}, Sandro Graffi,\footnote[2]{
Dipartimento di Matematica,  Universit\`{a} di Bologna, I- 40127 
Bologna (Italy) 
 (graffi@dm.unibo.it)} 
Johannes Sj\"ostrand\footnote[3]{Centre de Math\'ematiques, 
\'Ecole 
Polytechnique, F-91190 Palaiseau Cedex (France) 
(johannes@math.polytechnique.fr)}, 
\end{center}
\begin{abstract}
\noindent
Criteria  are  formulated both for the  existence and  for the 
non-existence of complex eigenvalues for a class of non 
self-adjoint operators in Hilbert space invarariant under a 
particular discrete symmetry. Applications to the  PT-symmetric 
Schr\"odinger operators are discussed.
 
\end{abstract}
\vskip 1cm   
%
 
%

%
%
%
%

\begsection{Introduction and statement of the results}
\setcounter{equation}{0}%
\setcounter{theorem}{0}%
\setcounter{proposition}{0}%
\setcounter{lemma}{0}%
\setcounter{corollary}{0}%
\setcounter{definition}{0}%

The  
 Schr\"odinger operators invariant under the combined application 
of a reflection symmetry operator $P$ and of the (antilinear) 
complex conjugation operation $T$ are called PT-symmetric. A 
standard class of such operators has the form $H=H_0+iW$ where:
\begin{enumerate}
\item
  $H_0$ is a self-adjoint realization of $-\Delta+V$ on 
some 
Hilbert space $L^2(\Omega); \Omega\subset\R^n$, $n\geq 1$; $V$ and $W$ are
real multiplication operators.
\item
  $V$ are even and odd with respect to  $P$, respectively:  $PV=V$, 
 $PW=-W$. $P$ is 
the parity operation 
$$ 
(P \psi)(x)=\psi((-1)^{j_1}x_1,\ldots,(-1)^{j_n}x_n), 
\qquad\psi\in 
L^2
$$ 
where $j_i=0,1$; $j_i=1$ for at least one $1\leq i\leq n$;
\end{enumerate}
If $T$ is the involution defined by complex conjugation:
$\ds (T\psi)(x)=\overline{\psi}(x)$, one immediately checks that 
$(PT)H=H(PT)$.
\par\noindent
$PT-$symmetric quantum mechanics (see e.g. \cite{Ah},\cite{Be1},\cite{Be2},\cite{Be3},
\cite{Cn1},\cite{Cn2},\cite{Cn3}, \cite{Zn1},\cite{Zn2}) requires the 
reality of the spectrum of $PT-$symmetric operators, recently 
proved, for instance, for the one dimensional odd anharmonic 
oscillators \cite{Tateo}, \cite{Shin}. Imposing boundary conditions 
along complex directions, ho\-we\-ver,  examples of $PT-$ 
sym\-me\-tric 
operators with complex eigenvalues have been constructed 
\cite{DD}. 
 It is therefore an important issue in this context to determine 
whether or not the spectrum  of $PT$-symmetric \Sc\ operators with 
standard $L^2$ boundary conditions at infinity is real. We deal 
with this problem only in perturbation theory, but we will obtain
 criteria both for existence of complex eigenvalues (Theorem 1.1) 
and for the reality of the spectrum (Theorem 1.2), in even  greater generality 
than 
 the $PT$ symmetry. 

Let ${\cal H}$ be a Hilbert space with scalar product denoted 
$(x|y)$, and  
$H_0:{\cal H}\to {\cal H}$ be 
a closed operator with dense  domain ${\cal
D}\subset{\cal H}$. Let $H_1$ be an   operator in ${\cal H}$ 
with 
${\cal D}(H_1
)\supset {\cal D}$. This entails that $H_1$ is bounded relative to 
$H_0$, i.e. there exist $b>0$, $a>0$ such that $\|H_1\psi\|\leq 
b\|H_0\psi\|+a \|\psi\|$ $\forall\,\psi\in {\cal D}$. We can therefore define  on ${\cal 
D}$ the operator family $H_\ep:=H_\ep=H_0+\ep H_1$,   $\forall\ep\in\C$.
\par We assume the following symmetry properties: there exists a 
unitary 
involution $J :{\cal H}\to {\cal H}$ mapping ${\cal D}$ to ${\cal 
D}$,
such that
\be
\label{H1}
{
JH_0=H_0^\ast J,\quad JH_1=H_1^\ast J }
\ee
In other words, $J$ intertwines $H_0$ and $H_1$ with the 
corresponding adjoint operators. Note that:
\begin{enumerate}
\item The properties  $J^2=1$ 
(involution) and $J^\ast=J^{-1}$ (unitarity) entail $J^\ast=J$, 
i.e. self-adjointness of $J$;
\item 
 The properties (\ref{H1}) entail, if $\ep\in\R$,  
$JH_\ep=H_\ep^\ast J$; therefore the spectrum $\sigma(H_\ep)$ 
of $H_\ep$ is symmetric with respect to the real axis if $\ep\in\R$.
\item An example of $J$ is the parity operator $P$.
\end{enumerate}
Let $H_0$ admit a real isolated eigenvalue $\l_0$ of  
multiplicity $2$ (both algebraic and geometric, i.e. we assume absence of 
Jordan blocks). Let $e_1,e_2$ be linearly independent 
eigenvectors, and $E_{\l_0}$ the eigenspace spanned by $ e_1,e_2$. 
Clearly $JE_{\l_0}:=E_{\l_0}^\ast$ is the eigenspace of $H_0^\ast$ 
corresponding to the eigenvalue $\overline{\l}_0=\l_0$, 
and hence the bilinear form $(u^\ast|v), u^\ast\in E_{\l_0}^\ast, v\in E_{\l_0}$ 
is non degenerate. 
Therefore we can choose $e_1,e_2$ in $E_{\l_0}$ in such a way that, 
writing $u=u_1e_1+u_2e_2$, the quadratic form $Q(u,u)=(Ju|u)$ on $E_{\l_0}$ assumes 
the canonical form
\be
\label{can}
Q(u,u)=\tau_1u_1^2+\tau_2u_2^2, \quad \tau_1=\pm 1, \tau_2=\pm 1
\ee
Notice that if $e_1^\ast, e_2^\ast$ is the dual basis, then (\ref{can}) means
that $Je_j=\tau_je^\ast_j$. 

Under these circumstances we want to 
prove the following
\begin{theorem}
\label{th1} With  the above assumptions and notations, consider the 
operator family $H_\ep$ for $\ep\in\R$. Denote:
\be
\label{elm}
H_{11}=(H_1e_1|e_1), \quad H_{22}=(H_1e_2|e_2), \quad H_{12}=(H_1e_1|e_2)
\ee
Then $(e_1|H_1e_1)\in\R$, $(e_2|H_1e_2)\in\R$ and there exists 
$\ep^\ast 
>0$ such that, for  $|\ep|<\ep^\ast$:
\begin{itemize}
\item[(i)]If $\tau_1\cdot\tau_2=-1$, and
\be
\label{cond}
 4|H_{12}|^2>(H_{11}-H_{22})^2
\ee
$H_\ep$ has a pair of non real,  complex 
conjugate eigenvalues near $\l_0$;
\item[(ii)]
If $\tau_1\cdot\tau_2=1$ $H_\ep$ has a pair of real eigenvalues 
near $\l_0$.
\end{itemize}
\end{theorem}
{\bf Remarks}
\begin{enumerate}
\item The above theorem applies to the $PT$-symmetric operator 
family $H_\ep=H_0+i\ep W$, where $H_0$  and $iW=H_1$ are as above. 
Here $J=P$, and hence $PH_0=H_0P$, $P(i\ep W)=-(i\ep W)P=(i\ep 
W)^\ast P$ so that $JH_\ep=H_\ep^\ast J$.  
In that case Assumption (\ref{cond}) 
follows from the weaker assumption $H_{12}\neq 0$ 
 because the $P-$symmetry of $H_0$ and the $P-$antisymmetry of $W$ 
entail $H_{11}=H_{22}=0$. Indeed, we have $Pe_j=\tau_je_j$ and
$$
H_{jj}=(iWe_j|e_j)=(iPWe_j,Pe_j)=(-iWPe_j|Pe_j)=-(iWe_j|e_j)=-H_{jj}
$$ 
\item
The physical relevance of Theorem 1.1 is best illustrated by an 
elementary e\-xam\-ple. Let ${\cal H}=L^2(\R^2)$  and  
$H_0:{\cal H}\to {\cal H}$ be 
the (self-adjoint)  two dimensional harmonic 
oscillator with frequencies $\om_1, \om_2$:
$$
H_0u=-\frac12\Delta u+\frac12(\om_1^2x_1^2+\om_2^2x_2^2)u
$$
We have 
$\ds \sigma(H_0)=\{E_{k_1,k_2}\}:=\{k_1\om_1+k_2\om_2+\frac{\om_1}{2}+\frac{\om_2}{2}\}, 
k_i=0,1,2\ldots, i=1,2$. Let again $H_\epsilon =H_0+i\epsilon W$, 
$\epsilon \in\R$, with
$$
W(x)=\frac{x_1^2x_2}{1+x_1^2+x_2^2}
$$
Then $W$ is bounded relative to $H_0$, and $PW=-W$ if 
$Pu(x_1,x_2)=u(x_1,-x_2)$ or $Pu(x_1,x_2)=u(-x_1,-x_2)$. Set 
$\om_1=1, \om_2=2$, $k_1=2, k_2=0$; i.e., we consider the 
eigenvalue $E_{2,0}=E_{0,1}$. Then for $|\ep|>0$ small enough $H_\ep$ has 
a 
pair of complex conjugate eigenvalues near $E_{2,0}$.

To see this, remark that 
$E_{2,0}=E_2(\om_1)+E_0(\om_2)=E_0(\om_1)+E_1(\om_2)$, where 
$\l_i(\om_i)=(k+1/2)\om_i$ are the eigenvalues of the 
one-dimensional harmonic oscillators with frequencies $\om_i, 
i=1,2$. $E_{2,0}$ has multiplicity $2$. A basis of eigenfunctions 
is given by
$$
\psi_1(x_1,x_2)=e_2(x_1)f_0(x_2); \qquad  
\psi_2(x_1,x_2)=e_0(x_1)f_1(x_2)
$$
Here $e_0$, $e_2$ are the eigefunctions corresponding to $E_0(1)$ 
and $E_2(1)$, respectively;  $f_0$, $f_1$ are the eigenfunctions 
corresponding to $E_0(2)$ and $E_1(2)$, respectively; note that 
$e_0$, $e_2$ and $f_0$ are 
even while $f_1$ is odd. To first order perturbation theory, the 
two eigenvalues $\Lambda_{j}(\ep): j=1,2$ of $H_\ep$ near $E_{2,0}$ are given 
by
$$
\Lambda_{j}(\ep)=E_{2,0}+i\ep \lambda_{j}
$$
where $\lambda_{j}: j=1,2$ are the eigenvalues of the $2\times 2$ 
matrix 
$$
W_{l,k}=\left(\begin{array}{ll} 
(W\psi_1|\psi_1)  & (W\psi_1|\psi_2)
\\
(W\psi_2|\psi_1)  & (W\psi_2|\psi_2)
\end{array}\right)
$$
Now $\psi_1$ is even, $\psi_2$ is odd, and $W$ is odd. Therefore $\tau_1\cdot \tau_2=-1$. Moreover: 
$(W\psi_1|\psi_1) = (W\psi_2|\psi_2)=0$, 
$(W\psi_2|\psi_1) = (W\psi_1|\psi_2) :=w>0$ 
Therefore $\lambda_{j}=\pm w$ and $\Lambda_{j}(\ep)=E_{2,0}\pm i\ep 
w$.  Hence the conditions of Theorem 1.1 (i) are satisfied and for $\ep$ small enough 
$H_\ep$ has a pair complex conjugate eigenvalues near $E_{2,0}$. 
\item
By essentially the same proof, the result
 of Theorem \ref{th1} remains true under the following more 
general 
conditions: under the above assumptions on $H_0$ and $H_1$  let 
$H_0$ admit two real, simple eigenvalues $E_1,E_2$. Let $d:=E_2-E_1$ be 
their relative distance; 
$D:={\rm dist}[(\sigma(H_0)\setminus\{E_2,E_1\}),\{E_2,E_1\}]$ 
their distance from the rest of the spectrum; $e_1, e_2$ 
the corresponding eigenvectors, all other notation being the same.   
Then if  $d/D$ is small enough the same conclusion of Theorem 
\ref{th1} holds provided $\ds |\ep H_{12}|>\frac{d}{2D}$.
\item Example: {\it Odd perturbations of quantum mechanical double 
wells: existence of complex eigenvalues.} 
\par\noindent
Let ${\cal H}=L^2(\R)$, 
$\ds H_0(\hbar)=-\hbar^2\frac{d^2}{dx^2}+x^2(1+x)^2$, $\ds 
D(H_0)=H^2(\R)\cap L^2_4(\R)$, $W(x)\in 
L^\infty_{loc}(\R)$, $|W(x)|\leq Ax^4$, $|x|\to\infty$, $W(1-x)=-W(x)$. Here 
$L^2_4(\R)=\{u\in L^2(\R)\,|\,x^4u\in L^2(\R)\}$. In this case it 
is known that $W$ is bounded relative to $H_0$; moreover 
$\ds d={\cal O}(e^{-1/c\hbar})$, $D={\cal O}(\hbar)$, $w={\cal 
O}(1)$  if $E_1,E_2$ are the two 
lowest 
eigenvalues, $\psi_1,\psi_2$ the corresponding eigenvectors and $w$ is defined as in
Point 2 above. 
Hence the conditions of Theorem 1 are fulfilled in the 
semiclassical regime provided $W$ is continuous at zero 
with  $W(0)\neq 0$ and that $|(e_1|We_2)|\geq 1/C$ and thus there exist $A>0, B>0, C>0$ such 
that
$H_\ep(\hbar):=H_0+i\ep W$ will have at least a 
pair of complex conjugate eigenvalues for $\ds Ae^{-B/\hbar^2}<\ep  
w<< C\hbar$. Equivalently,
we may consider the double well family $\ds 
H_0(g)=-\frac{d^2}{dx^2}+x^2(1+gx)^2$ defined on the same domain. 
Here $\ds d={\cal O}(e^{-1/g^2})$, $D={\cal O}(1)$, $w={\cal 
O}(1)$. 
The same argument holds for the general case $H_0=-\hbar^2\Delta 
+V(x)$, where $V:\R^n\to\R$ is smooth, has two equal quadratic  
minima and diverges positively as $|x|\to\infty$; $W(x)\in 
L^\infty_{loc}(\R^n)$, $|W(x)|\leq A V(x)$ as $|x|\to\infty$ 
because the estimate for $d$ is the same as above\cite{HS}.
\end{enumerate}
The second result concerns the opposite situation, a criterion ensuring the reality of the spectrum. In this case the natural assumption is the simplicity of the spectrum of $H_0$ in addition to its reality. Therefore for the sake of simplicity we assume $H_0$ self-adjoint.
\begin{theorem}
\label{th2}
Let the self-adjoint operator $H_0$ be bounded below (without loss of generality, 
positive), and let $H_1$  be continuous. Let $H_0$ have discrete spectrum, $\sigma(H_0)=\{0\leq 
\l_0 
<\l_1 \ldots < \l_l< \ldots \}$, with the property
\be
\label{dist}
\delta:=\inf_{j\geq 0}\;[\l_{j+1}-\l_j]/2 >0.
\ee
Assume that all eigenvalues are simple.
Then $\sigma(H(\ep))\in\R$ if $\ep\in\R$, $\ds 
|\ep|<\frac{\delta}{\|H_1\|}$.
\end{theorem}
\vskip 0.2cm\noindent
{\bf Example}
\par\noindent
Here again ${\cal H}=L^2(\R)$; $\ds H_0=-\frac{d^2}{dx^2}+V(x)$, 
$\ds D(H_0)=H^2(\R)\cap D(V)$. $V(x)=kx^{2m}$, $k>0$, $m\geq 1$; $W(x)\in 
L^\infty(\R)$, $W(-x)=-W(x)$. We have: $\sigma(H_0)=\{\l_l\}, 
n=0,1,\ldots$; 
$$
 \l_n \sim  k^{\frac{1}{2m}}n^{\frac{2m}{m+1}}, \quad n\to\infty
$$
Each eigenvalue 
$\l_n$ is simple. Clearly $\ds \delta\geq 1$. Denote now 
$H_\ep:=H_0+i\ep W$ the operator family in $L^2(\R)$ defined by 
$\ds H_\ep= H_0+H_1$, $H_1=i\ep W$, $D(H_\ep)=D(H_0)$. Then 
$H_\ep$ 
has real discrete spectrum for $|\ep|<\|W\|_\infty^{-1}$.
\vskip 1.5cm\noindent
\section{Proof of the results}
\setcounter{equation}{0}%
\setcounter{theorem}{0}%
\setcounter{proposition}{0}%
\setcounter{lemma}{0}%
\setcounter{corollary}{0}%
\setcounter{definition}{0}%
{\bf Proof of Theorem 1.1}
\par\noindent
The proof is based on perturbation theory and consists in two 
steps. In the first one we show that the $2\times 2$ matrix 
generated by restricting the perturbation $H_1$ to $E_{\l_0}$ is 
antihermitian in case (i) of Theorem 1.1 and  Hermitian in case 
(ii). 
In the second step we show by the method of the Grushin reduction (see, e.g.\cite{HS1})  
that for $\ep$ suitably small the control of the above $2\times 2$ 
matrix is enough to establish the result.

Let $\{e_1,e_1\}$ be once more a basis  in $E_{\l_0}$ such that (1.2) holds, and denote by 
$e_1^\ast,e_2^\ast$ the dual basis in the dual 
subspace $E_{\l_0}^\ast=JE_{\l_0}$. Clearly 
$Je_j=\tau_j e_j^\ast$, $\tau_j=\pm 1$. We denote $\Pi_0$
the spectral projection from ${\cal H}$ to $E_{\l_0}$. Explicitly:
\be
\label{Pi}
\Pi_0u=(u|e_1^\ast)e_1+(u|e_2^\ast)e_2
\ee
Consider now the rank $2$ operator family $\Pi_0 H_\ep \Pi_0$ acting on $E_{\l_0}$. The representing 
$2\times 2$ matrix is:
\be
\label{matrice}
H(\ep)_{j,k}=\l_0I+\ep H^1_{j,k}, \quad 
H^1_{j,k}=(H_1e_k|e^\ast_j),\;j,k=1,2
\ee
Now $JH_0=H_0^\ast J$,  $J\Pi_0=\Pi_0^\ast J$. We also have $JH_1=H_1^\ast J$. Therefore:
$$
(JH_1e_k|e_j)=(H_1e_k|Je_j)=\tau_j(H_1e_k|e^\ast_j)=\tau_jH^1_{j,k}
$$
and in the same way
$$
(JH_1e_k|e_j)=(H_1^\ast 
Je_k|e_j)=(Je_k|H_1e_j)=\tau_k(e_k^\ast|H^1e_j)=\tau_k
v\overline{(H_1e_j|e^\ast_k)}=\tau_k\overline{H^1_{k,j}}
$$
Summing up:
$$
\tau_jH^1_{j,k}=\tau_k\overline{H^1_{k,j}}
$$
Therefore, if $\tau_1\tau_2=1$ the matrix $H(\ep)_{j,k}$ is 
hermitian for $\ep\in\R$ and its eigenvalues are real; if instead 
$\tau_1\tau_2=-1$ the matrix $H(\ep)_{j,k}$ has a real diagonal part
and an  antihermitian off diagonal part  for 
$\ep\in\R$ and its eigenvalues are complex conjugate. This 
completes the first step. 
\newline
We  want now to construct an approximate inverse of $H_\ep-z$ near 
$\l_0$ by solving a Grushin problem. In this context it is 
equivalent to the Feshbach reduction, and provides a convenient 
formalism for it. To this end, define the 
operators $R_+, R_-$, ${\cal P}_0(z)$ in the following way:
\begin{eqnarray}
\label{R} 
R_+:{\cal H}\to{\C}^2, \;R_+u(j)=(u\vert e_j^\ast),\; j=1,2;
\\
R_-:{\C}^2\to{\cal H}, \; R_-u_-=\sum_{j=1}^2u_-(j)e_j,\qquad
\\
\label{2.4}
{{\cal P}_0(z)=\pmatrix{ H_0-z &R_-\cr R_+ &0}:{\cal D}\times 
{\C}^2\to
{\cal H}\times {\C}^2.}
\end{eqnarray}
Note that we have identified $E_{\l_0}$ with its representative 
$\C^2$, and that  $R_+R_-=I$, the $2\times 2$ identity matrix.
\par\noindent
The associated Grushin system is
\be
\label{Gr}
\left\{\begin{array}{l} (H_0-z)u+R_-u_-=f
\\
R_+u =f_+
\end{array}\right.
\ee
where $u\in{\cal D},f\in{\cal H}$, $u_-,f_+\in\C^2$. $z\in\C$ 
belongs to a neighborhood of $\l_0$ at a positive distance from $\sigma(H_0)\setminus\{\l_0\}$. After 
determining $u_-$ in such a 
way that $f-R_-u_-\in (1-\Pi_0){\cal H}$ the first 
equation 
can be solved  for $u(z)\in (1-\Pi_0){\cal H}$ and hence the 
problem is 
reduced to the the rank $2$ equation $R_+u(z)=f$. 
To solve explicitly, remark that,  for  every $z$ in the complex 
complement of 
$\sigma (H_0)\setminus\{
\l_0\}$, ${\cal P}_0(z)$ has the \bdd{} inverse,
\be
\label{2.5}
{
{\cal E}_0(z)=\pmatrix{E^0(z)&E_+^0(z)\cr E_-^0(z) &E_{-+}^0(z)},
}
\ee
with
\begin{eqnarray}
\label{2.6}
{E^0(z)=(H_0-z)^{-1}(1-\Pi ),\quad E_+^0(z)=R_-,}
\\
\nonumber
{E_-^0(z)=R_+,\quad E_{-+}^0(z)=(z-\l_0)I.}
\end{eqnarray}
where $I$ is the $2\times 2$ identity matrix.  The spectral problem within $E_{\l_0}$ 
is thus reduced to the inversion of $E_{-+}^0(z)$, and 
obviously its solution is represented by $\l_0, e_0, e_1$.
 \par Now restrict the attention to the  set of complex $z$ 
with 
${\rm
dist\,}(z,\{ \l_0\} )<1/2R$, where 
\be
\label{normaR}
R:=\|E^0(\l_0)\|=\|(1-\Pi_0)(H_0-\l_0)^{-1}\|
\ee
so that by the geometrical series expansion
\be
\label{normaE}
\|E^0(z)\|\leq \frac{R}{1-|z-\l_0|R}
\ee
Consider the operator from ${\cal D}\times \C^2$ to ${\cal H}$ 
defined as 
\be
\label{2.8}
{
{\cal P}_\epsilon (z)=\pmatrix{H_\epsilon -z &R_- \cr R_+  
&0}.
}
\ee
associated to the Grushin system
\be
\label{Gr!}
\left\{\begin{array}{l} (H_\ep-z)u+R_- u_-=f
\\
R_+u=f_+
\end{array}\right. .
\ee
Then
\be
\label{2.9}   
{{\cal P}_\epsilon (z){\cal E}_0(z)=1+\pmatrix{i\epsilon H_1E^0(z) 
&i\epsilon
H_1E_+^0(z)\cr 0 &0}=:1+{\cal K}.}
\ee
It is routine to check that ${\cal P}_\epsilon (z)$ has the 
inverse
\be
\label{2.10}
{{\cal E}_\epsilon (z)=\pmatrix{E^\epsilon (z) &E_+^\epsilon 
(z)\cr
E_-^\epsilon (z) &E_{-+}^\epsilon (z)},}
\ee
with
\begin{eqnarray}
\label{2.10bis}
E^\ep(z)&=&\sum_{n=0}^\infty  ({\epsilon \over
i})^nE^0(H_1E^0)^n, 
\\
 E^\ep_+(z)&=&\sum_{n=0}^\infty  ({\epsilon 
\over
i})^n(E^0H_1)^nE_+^0 
\\
\label{2.10ter}
E^\ep_-(z)&=&\sum_{n=0}^\infty  ({\epsilon \over 
i})^nE_-^0(H_1E^0)^n,
\\
\label{2.10quater}
E^\ep_{-+}(z)&=&E_{-+}^0+\sum_{n=1}^\infty ({\epsilon \over
i})^nE_-^0(H_1E^0)^{n-1}H_1E_+^0. 
\end{eqnarray}
where all the series will be proved to have a positive convergence 
radius (convergence means here uniform, or, equivalently, in the 
norm operator sense). We also recall the well known fact that $z$ is an eigenvalue
of $H_\ep$ precisely when ${\rm det}\,E^\ep_{-+}(z)=0$.  
\par We next derive the appropriate symmetries for the inverse 
\op{}s \cite{HS1}. From $JH_\ep=H¬_\ep^\ast J$ we get:
\begin{eqnarray*}
JR_-u_-&=&\sum_{j=1}^2 u_-(j)J
e_j = \sum_{j=1}^2 (\tau 
u_-)(j)e_j^\ast, \quad \tau:=\left(\begin{array}{ll} \tau_1 & 0
\\ 0 & \tau_2\end{array}\right)
\\
R_+^\ast u_-&=&\sum_{j=1}^2 u_-(j)e_j^\ast
\end{eqnarray*}
where the second equation follows from 
$$ 
(R_+u|u_-)=\sum_{j=1}^2 \overline{u_-(j)}(u|e_j^\ast)
$$
We thus conclude:
$$
JR_-u_-=R_+^\ast\tau u_-,\quad R_-^\ast J=\tau R_+
$$
Therefore:
\begin{eqnarray*}
\left(\begin{array}{ll} J
 & 0
\\ 0 & \tau\end{array}\right)
\left(\begin{array}{ll} H_\ep -z & R_-
\\ R_+ & 0\end{array}\right)=
\left(\begin{array}{ll} J(H_\ep -z) & JR_-
\\ \tau R_+ & 0\end{array}\right)
\\
=
\left(\begin{array}{ll} (H_\ep^\ast -z)J & R_+^\ast\tau
\\  R_-^\ast J & 0\end{array}\right)=
\left(\begin{array}{ll} (H_\ep^\ast -z) & R_+^\ast
\\  R_-^\ast  & 0\end{array}\right)\left(\begin{array}{ll} J
 & 0
\\ 0 & \tau\end{array}\right)
\end{eqnarray*}
whence
\be
\label{17}
\left(\begin{array}{ll} J
 & 0
\\ 0 & \tau\end{array}\right){\cal P}_\ep(z)={\cal 
P}_\ep(\overline{z})^\ast\left(\begin{array}{ll} J
 & 0
\\ 0 & \tau\end{array}\right)
\ee
Since ${\cal E}(z)={\cal P}(z)^{-1}$, taking right and left 
inverses we get 
$$
{\cal E}(\overline{z})^\ast\left(\begin{array}{ll} J
 & 0
\\ 0 & \tau\end{array}\right)=\left(\begin{array}{ll} J
 & 0
\\ 0 & \tau\end{array}\right){\cal E}(z)
$$
that is
\be
\label{18}
\left(\begin{array}{ll} E(\overline{z})^\ast & 
E_-(\overline{z})^\ast
\\  E_+(\overline{z})^\ast & 
E_{-+}(\overline{z})^\ast\end{array}\right)\left(\begin{array}{ll} 
J
 & 0
\\ 0 & \tau\end{array}\right)=\left(\begin{array}{ll} J
 & 0
\\ 0 & \tau\end{array}\right)
\left(\begin{array}{ll} E({z}) & E_+({z})
\\  E_-({z}) & E_{-+}({z})\end{array}\right)
\ee
In particular:
$$
E_{-+}(\overline{z})^\ast\tau=\tau E_{-+}({z})
$$
We can thus conclude that, for $z\in\R$, if $\tau_1\cdot\tau_2=1$ the  $2\times 
2$ 
matrix $E_{-+}({z})$ is Hermitian, and antihermitian off the diagonal with real diagonal elements if 
if $\tau_1\cdot\tau_2=-1$.
\newline
It remains to be proved the norm convergence of the expansions 
(\ref{2.10bis},\ref{2.10ter},\ref{2.10quater}). We have, by the 
relative boundedness condition $\|H_1\psi\|\leq 
b\|H_0\psi\|+a\|\psi\|$ and (\ref{normaE}):
\begin{eqnarray*}
\|H^1E^0\|&=&\|H^1(H_0-z)^{-1}(1-\Pi_0)\|\leq 
\\
&\leq& b\|H_0(H_0-z)^{-1}(1-\Pi_0)\|+a\|(H_0-z)^{-1}(1-\Pi_0)\|
\\
&\leq & b\|(H_0-z)(H_0-z)^{-1}(1-\Pi_0)\|+
\\
&+& b|z|\|(H_0-z)^{-1}(1-\Pi_0)\|+a\|(H_0-z)^{-1}(1-\Pi_0)\|
\\
&\leq& b\|1-\Pi_0\|+\frac{(b|z|+a)R}{1-|z-\l_0|R}<K
\end{eqnarray*}
for some $K(z)>0$ because $|z|<R/2$. Therefore
\begin{eqnarray*}
\|E^0(H^1E^0)^n\|\leq K^{n+1}, \quad \|(E^0H^1)^nE^0_+\|\leq 
K^{n+1}, 
\\
\|E^0_-(H^1E^0)^n\|\leq K^{n+1}, \quad 
\|E^0_-(H^1E^0)^{n-1}H_1E^0_+\|\leq K^{n+1}
\end{eqnarray*}
Hence the expansions 
(\ref{2.10bis},\ref{2.10ter},\ref{2.10quater}) 
are norm convergent. 
\par\noindent To conclude the proof we have to verify that the first order 
truncation of the expansion for $E_{_+}(z)$ yields nonreal eigenvalues, 
and that the higher order terms can be neglected. To this end, 
first remark that without loss of generality we may assume $\l_0=0$. Then the expansion 
(\ref{2.10quater}) yields:
$$
-E^\ep_{-+}(z)=\left(\begin{array}{ll}
\ep H_{11}-z & \ep H_{12}
\\
-\ep \overline{H}_{12} & \ep H_{22}-z
\end{array}\right) +O(\ep^2)
$$ 
uniformly with respect to $z$, $|z|<1/2R$. Therefore
\begin{eqnarray*}
{\rm det}E^\ep_{-+}(z)=z^2-(H_{11}+H_{22})\ep z+
\ep^2(|H_{12}|^2+H_{11}H_{22})+O(\ep^3+\ep^2|z|)=
\\
=[z-\ep (H_{11}+H_{22})/2]^2+\ep^2[|H_{12}|^2-
(H_{11}-H_{22})^2/4] +O(\ep^3+\ep^2|z|)
\end{eqnarray*}
Now ${\rm det}E^\ep_{-+}(z)$, which is real for $z\in\R$,  clearly has no zeros for 
$z\in\C$, $\ep <<|z|<<1$. On the other hand, for $z=O(\ep)$, i.e. $z=\ep w$, $w=O(1)$, 
\begin{eqnarray*}
{\rm det}E^\ep_{-+}(z)&=&\ep^2\{[w- (H_{11}+H_{22})/2]^2+|H_{12}|^2-
(H_{11}-H_{22})^2/4\}
\\
&+& O(\ep^3(1+O(1))
\end{eqnarray*}
Therefore if $4|H_{12}|^2>(H_{11}-H_{22})^2$ there cannot be real zeros for $\ep$ 
suitably small. We can thus conclude that 
${\rm det}E^\ep_{-+}(z)$ is zero for $z=\Lambda_\pm(\ep)$,
$$
\Lambda_\pm(\ep)=\frac12[{H_{11}+H_{22}}\pm i\ep \sqrt
{4|H_{12}|^2-(H_{11}-H_{22})^2}]+O(\ep^2)
$$
and this concludes the proof of the Theorem.
\vskip 1.5cm\noindent
{\bf Proof of Theorem \ref{th2}}

Let us first recall that under the present assumptions $H_\ep$ is 
a type-A 
holomorphic family of operators in the sense of Kato (see \cite{Ka}, Chapter VII.2) with compact resolvents 
$\forall\,\ep\in\C$. Hence  $\sigma(H_\ep)=\{\l_l(\ep)\}: l=0,1,\ldots$. 
In particular:
\begin{itemize}
\item[(i)] the \ev s $\l_l(\ep)$ are locally holomorphic functions 
of $\ep$ with only algebraic singularities; 
\item[(ii)]
the \ev s $\l_l(\ep)$ are stable, namely given any eigenvalue 
$\l(\ep_0)$ of $H_{\ep_0}$ there is 
exactly  one \ev\ $\l(\ep)$ of $H_\ep$  such that $\ds 
\lim_{\ep\to\ep_0}\l(\ep)=\l(\ep_0)$;
\item[(iii)] the Rayleigh-\Sc\ perturbation expansion for the 
eigenprojections and the eigenvalues near any 
eigenvalue $\l_l$ of $H_0$ has convergence radius $\ds 
\delta_l/\|H_1\|$ where $\delta_l$ is half the isolation distance 
of $\l_l$. 
\end{itemize}
Remark that since $\delta_l\geq \delta\;\forall\,l$, 
all the series will be convergent for all $\ep\in \Om_{r_0}$; 
$\Om_{r_0}:=\{\ep\in\C: |\ep|< r_0\}$, where 
$r_0:=\delta/\|H_1\|$ is a uniform lower bound for all convergence 
radii. 
\par\noindent
Assume now  without loss of generality, to simplify the notation, 
$\|H_1\|=1$. By hypothesis $|\l_l-\l_{l+1}|\geq 
2\delta>0\,\forall\,l\in\N$. 
First remark that 
if $\ep\in\R$, $|\ep|<r_0$ and $\l(\ep)$ is an eigenvalue of 
$H_\ep$ 
then $|\Im \l(\ep)|<\delta$, i.e. $\sigma(H_\ep)\cap \C_\delta=\emptyset$, 
$\C_\delta:=\{z\in\C\,|\,|\Im\,z|\geq\delta\}$. Set indeed 
$$
R_0(z):=[H_0-z]^{-1},\quad z\notin\sigma(H_0)
$$
Then $\forall\,z\in\C$ such that $|\Im z|\geq \delta$ we have
\be
\label{3.77}
\|\ep H_1R_0(z)\|\leq 
|\ep|\cdot\|H_1\|\cdot\|R_0(z)\|\leq\frac{|\ep|}{{\rm 
dist}[z,\sigma(H_0)]}\leq\frac{|\ep|}{|\Im z|}
\ee
Hence the resolvent
$$
R_\ep(z):=[H_\ep-z]^{-1}=R_0(z)[1+\ep H_1R_0(z)]^{-1}
$$
exists and is bounded if $|\Im z|\geq \delta$ because (\ref{3.77}) entails the  
uniform norm convergence of the Neumann expansion for the resolvent:
\begin{eqnarray*}
\|R_\ep(z)\|=\|[H_\ep-z]^{-1}\|=\|R_0(z)\sum_{k=0}^\infty[-\ep H_1R_0(z)]^{k}\|\leq
\\
\leq  \|R_0(z)\|\sum_{k=0}^\infty|\ep^k|\|H_1R_0(z)]\|^{k}
\leq \frac{|\ep|}{|\Im z|-\ep}
\end{eqnarray*}
  Now $\forall\,l\in\N$ let 
$Q_l(\delta)$ denote the open square of side $2\delta$ centered at $\l_l$.
Since $|\l_l-\l_{l+1}|\geq 2\delta$, it follows as in  (\ref{3.77}) 
that 
$R_\ep(z)$ exists and is bounded for $z\in\partial Q_l(\delta)$, 
the boundary of $Q_l(\delta)$. We can therefore, according to the standard procedure 
(see e.g.\cite{Ka}, Chapter III.2) define 
the strong Riemann integrals 
$$
P_l(\ep)=\frac{1}{2\pi i}\int_{\partial Q_l(\delta)}R_\ep(z)\,dz, \qquad l=1,2,\ldots
$$
As is well known, $P_l$ is the spectral projection onto the part of $\sigma(H_\ep)$ 
inside $Q_l$. Since $H_\ep$ is a holomorphic family in $\ep$, by well known results 
(see e.g. \cite{Ka}, Thm. 
VII.2.1),  
the same is true for $P_l(\ep)$ for all $l\in\N$. In particular this entails 
the continuity of $P_l(\ep)$ for $|\ep|<r_0$. Now  $P_l(0)$ is a one-dimensional: 
hence 
the same is true for $P_l(\ep)$. As a consequence, there is one and only one point of 
$\sigma(H_\ep)$ inside any $Q_l$. Now  $\sigma(H_\ep)$ is discrete, and thus any 
such point is an eigenvalue; moreover, any such point is real 
for $\ep$ real because $\sigma(H_\ep)$ is symmetric with respect to the real axis. 
Finally, we note that 
if $z\in\R$, $\ds z\notin \bigcup_{l=1}^\infty ]\l_l-\delta,\l_l+\delta[$  the 
Neumann series (\ref{3.77}) is convergent and the resolvent  $R_\ep(z)$ is 
there continuous. This concludes the proof of Theorem 1.2. 

\vskip 1.5cm\noindent

\end{document}